# Critical Assessment of Mass and Lattice Disorder in Thermal Conductivity Prediction for Medium and High Entropy Ceramics

Yuxuan Wang[a], Guoqiang Lan[a], Jun Song[a,*]

*a Department of Mining and Materials Engineering, McGill, University, Montreal, Quebec, H3A 0C5, Canada*

** Corresponding author. E-mail address: jun.song2@mcgill.ca (J. Song).*

## Abstract

Medium and high entropy ceramics, with their distinctive disordered structures, exhibit ultra-low thermal conductivity and high temperature stability. These properties make them strong contenders for next generation thermal barrier coating (TBC) materials. However, predicting their thermal conductivity has been challenging, primarily due to their unique phonon scattering mechanisms. Apart from the conventional phonon-phonon scattering mechanism, the phonon-disorder scattering, comprising both mass and force constant disorder, are also expected to make significant contribution in determining the thermal conductivity of medium and high entropy ceramics. However, it remains challenging to quantify the phonon-disorder contribution, particular in the aspect of force constant disorder. Here we demonstrated a relationship between the lattice disorder, a quantity more readily calculable, with force constant disorder, rending it possible to substitute the force constant disorder by lattice disorder. Based on this relationship and drawing inspiration from Klement's equation of static imperfection, we have developed a model that quantitatively assess the connection between disorder and thermal conductivity. Applying our model to the medium/high entropy rocksalt and pyrochlore oxides as the representative, we found good alignment between the theoretical predictions and experimental measurements of thermal conductivities, confirming the validity of our model. The model developed offers a critical predictive tool for rapid screening of TBC materials based on medium and high entropy ceramics.

# 1. Introduction

High entropy ceramics have been wildly concerned because of their unique thermal properties. According to Gibbs free energy expression: $G = H - TS$, high entropy will accelerate the reduction of total energy especially at high temperature, which encourage the phase stabilization [1-4]. What's more, mass disorder and force constant disorder in high entropy ceramics strengthen the phonon scattering mechanism and further reduce the thermal conductivity[5-7]. These two characters make high entropy ceramics promising candidate for next generation thermal barrier coating materials. Besides, in some medium entropy ceramics, excellent thermal properties were also found[8-11]. The difficulty of developing new medium/high entropy ceramics mainly comes from materials screening because permutation of elements choice in medium/high entropy ceramics is enormous. Traditional experimental measurement is time consuming and materials wasting. And for computational predictions, models of thermal conductivity based on phonon-phonon scattering mechanisms have been developing for decades and could accurately predict most of the simple crystal structure[12-14]. However, for medium/high entropy materials that have unique structure, there still lack effective thermal conductivity prediction models. The thermal conductivity of medium/high entropy ceramics comes from both phonon-phonon and phonon-disorder scattering, here disorder can be regard as a certain kind of quasi-particle that combines the effects from mass and force constant disorder [15-17]. The current method to calculate thermal conductivity of high entropy material includes VCA[18] (virtual crystal approximation), CPA[19] (coherent potential approximation) and SPU[20] (supercell phonon unfolding). Both VCA and CPA are mean field theory, which average the difference of mass and/or force constant effect and cannot reflect the influence from local disorder environment. And for recently developed SPU method, which can unfold the phonon dispersion information of supercell onto that of primitive cell and make the disorder system to be comparable with ordered structure, phonon-phonon (anharmonicity) scattering contribution is not considered in. What's worse, SPU method requires a huge amount of calculation resource to solve large amount of independent phonon dispersion curve.

Wright etc. studied the effect of atomic size disorder on thermal conductivity of 22 pyrochlore high entropy ceramics and found they have nearly negative linear relationship. However, these atomic sizes comes from empirical atomic size and cannot describe the real situation in different composition[21]. Meanwhile, other researchers found that lattice distortion (or atomic displacement) might be a more effective indicator of many properties like strength[22], thermal [23] and electrical[24] properties. And significantly, lattice distortion can be readily calculated through some ab-initio methods and can relatively accurately descript the ad-hoc disorder information in materials.

Therefore, in this paper, based on mass disorder and lattice disorder, we are trying to build a comprehensive quantitative prediction model for thermal conductivity of high entropy ceramic material. Inspired by Klements' thermal conductivity model of static imperfection, thermal conductivity is inversely proportional to degree of mass and force constant disorder. Firstly, we built root square relationship between lattice disorder, which is more general and easy-obtained, and complicated force constant disorder and then use lattice disorder to replace force constant disorder in the model. Subsequentially, prediction model was examined in two kind of medium/high entropy system, rocksalt oxides and pyrochlore oxides. Results indicate that our prediction model works well in these two kind of medium/high entropy system and reach very high coefficient of determination, which proved the feasibility of our model. Details and discussion will be provided in the following chapters.

# 2. Methodology

## 2.1 Density functional theory and supercell construction

Supercell volume optimization and atomic relaxation were calculated by DFT (density functional theory), performed using Vienna ab initio simulation package (VASP). For pseudopotential treatment, Perdew - Burke - Ernzerhof (PBE) GGA approach based on plane-wave base sets was adopted. The electron - core interaction was described by the Bloch's projector augmented wave method (PAW) within the frozen-core approximation. Valence configuration of different elements in rocksalt structure are as follow: Mg - $2p^6$ $3s^2$, Zn - $4s^2$ $3d^{10}$, Co - $4s^1$ $3d^8$, Cu - $3p^6$ $4s^1$ $3d^{10}$, Ni - $3p^6$ $4s^1$ $3d^9$, Sc - $3s^2$ $3p^6$ $4s^2$ $3d^1$, Sb - $5s^2$ $5p^3$, Sn - $5s^2$ $4d^{10}$ $5p^2$, Cr - $3p^6$ $4s^1$ $3d^5$, Ge - $4s^2$ $3d^{10}$ $4p^2$ and O - $2s^2$ $2p^4$. The value of Hubbard energy for 3*d* orbit of Cr, Co and Ni were set following Jain's research[25]. And for pyrochlore structure, the valence configuration of elements are: La - $5s^2$ $5p^6$ $6s^2$ $5d^1$, Pr - $5s^2$ $5p^6$ $6s^2$ $5d^1$, Nd - $5s^2$ $5p^6$ $6s^2$ $5d^1$, Sm - $5s^2$ $5p^6$ $6s^2$ $5d^1$, Eu - $5s^2$ $5p^6$ $6s^2$ $5d^1$, Gd - $5s^2$ $5p^6$ $6s^2$ $5d^1$, Yb - $5s^2$ $5p^6$ $6s^2$ $5d^1$, Ti - $3p^6$ $4s^1$ $3d^3$, Zr - $4s^2$ $4p^6$ $5s^2$ $4d^2$, Sn - $4s^2$ $3d^{10}$ $4p^2$, Hf - $5p^6$ $6s^2$ $5d^2$, O - $2s^2$ $2p^4$.

The supercell construction was realized by SQS (special quasi-random supercells) method through Python package *ICET* [26]. Only the positions of metal atoms were randomly replaced by different equiatomic dopants while keeping unchanged of oxygen positions. In rock salt structure, for 5-cation oxides, primitive cell was used to extend to its 5×5×5 supercell (250 atoms) while conventional cell was used to extend to its 3×3×3 supercell (216 atoms) for other compositions. For pyrochlore structure, primitive cell of $La_2Zr_2O_7$ was used to extend to its 3×3×3 supercell (594 atoms). And then, different metal elements will randomly replace La and Zr sites according to the composition design referred by Wright's study through SQS method. The composition design of rocksalt and pyrochlore structure medium and high entropy oxides are shown in Table1 and Table 2 of appendix part.

And for first step volume optimization, cell shape and atomic positions were firstly fixed and the lattice constants with lowest energy were obtained. The energy convergence for electron steps and ion steps were set as $5\times10^{-7}$ eV/atom and $5\times10^{-6}$ eV/atom respectively. And then lattice constants were fixed to just relax the atom positions. Gamma-centered k-mesh of 1 × 1 × 1 for Brillouin-zone integrations and cutoff energy of 520 eV for the plane-wave-basis were used in DFT calculation with convergence criteria for energy of electron and ion steps on each ion set as $5 \times 10^{-8}$ eV and $5 \times 10^{-8}$ eV. The real-space force constants in rock salt structure were calculated with finite displacement method through above built supercell.

## 2.2 SPU and thermal conductivity indicator

Supercell phonon-unfolding (SPU) method stems from a kind of band-unfolding treatment dealing with electron and phonons in random alloying system[27]. By unfolding the band of large random alloy supercell onto that of small primitive-cell periodic basis, an effective band spectra can be obtained. In this paper, we used the unfolding program *upho* written by Ikeda etc[28]. The unfolded spectral function is defined as: $A(\boldsymbol{k}_k, \omega) = \sum_J \left|\left[\hat{P}^{\boldsymbol{k}_k}\tilde{v}(\boldsymbol{K}, J)\right]_l\right|^2 \delta[\omega - \omega(\boldsymbol{K}, J)]$, where $\boldsymbol{k}_k$ is wave vector of primitive cell while $\boldsymbol{K}$ is from supercell, $J$ is band index and δ means delta function. In this expression, $\hat{P}^{\boldsymbol{k}_k}$ represents projection operator at $\boldsymbol{k}$ point in the primitive Brillouin zone, and $\tilde{v}(\boldsymbol{K}, J)$ denotes the eigenvectors of dynamical matrix of the supercell for wave vector $\boldsymbol{K}$. Delta function δ were smeared by the Lorentzian functions with half width at half maximum (HWHM) of 0.05THz for spectra plotting. And the linewidths of phonon band were defined as the full width at half maximum (FWHM) of unfolded spectra through Lorentzian function fitting. Referring Boltzmann equation with a relaxation time approximation to describe thermal conductivity[12], we built a thermal conductivity prediction indicator Λ of high entropy ceramics. Consider that phonon linewidth is inversely proportional to phonon lifetime, indicator $\Lambda$ can be written as: $\Lambda = \frac{A}{NV_0}\sum_{\lambda}\frac{Cv_{\lambda}\cdot v_{\lambda}{}^{2}{}_{x}}{lw_{\lambda}}$, where $A$ is adjustment parameter, $V_0$ is volume of primitive cell, $N$ is number of wave vector, $C_v$ is heat capacity, $v$ is phonon velocity, and $lw$ is phonon linewidth.

## 2.3 Prediction model construction

In Klement's theory of static imperfection on thermal conductivity, three kind of disorder contribute to the phonon lattice scattering: mass difference, elastic bond difference and elastic strain difference [29]. According to perturbation theory and isotropic continuum approximation, the effects of imperfection can be expressed as:

$$\tau_{\lambda}^{-1} = \frac{\Gamma_i V_0}{4\pi\bar{v}^3}\omega^4(\lambda) \tag{1}$$

where $\tau$ is phonon lifetime, $\lambda$ is independent vibration mode, $V_0$ is unit cell volume, $\omega$ is phonon frequency, $\bar{v}$ is average sound velocity while $\Gamma_i$ denote degree of point defect imperfection, which can be expressed as:

$$\Gamma_i = x_i\left[\left(\frac{\Delta M_i}{M}\right)^2 + 2\times\left(\left(\frac{\Delta G_i}{G}\right) - 2\times 3.2\gamma\left(\frac{\Delta\delta_i}{\delta}\right)\right)^2\right] \tag{2}$$

where $x_i$ is ratio of impurity, $M_i$ is the mass of impurity and $M$ is atomic mass of host lattice. $G_i$ is the average stiffness constant of the nearest neighbor bonds from impurity to host lattice and $G$ is the corresponding value host lattice. $\gamma$ is Gruneisen's parameter and $\delta_i$ is radius of impurity while $\delta$ is that of host matrix.

Klement's model is based on point-defect scattering by impurity and consider the nearest neighbor forces as harmonic bonding and acting on simple cubic lattice. However, in high entropy or medium entropy materials, there is no strictly-defined host atoms, all kinds of atoms are randomly distributed on certain crystal sites. Interatomic bonding and elastic stain filed will become more complex than that in simple solid solution. Braun etc.[7] examined the applicability of Clement's model in high entropy rocksalt structure oxides by curve fitting, where properties of host matrix were replaced by average value of every component and total disorder was calculated as the summation of every components, which shown as:

$$\Gamma_{tot} = \sum_{N} x_i\left[\left(\frac{\Delta M_i}{\bar{M}}\right)^2 + 2\times\left(\left(\frac{\Delta G_i}{\bar{G}}\right) - 2\times 3.2\gamma\left(\frac{\Delta\delta_i}{\bar{\delta}}\right)\right)^2\right] \tag{3}$$

Both the experimental data from binary alloy $Co_{0.2}Ni_{0.8}O$, $Zn_{0.4}Mg_{0.6}O$, $Co_{0.25}Ni_{0.75}O$ and high entropy J14($Mg_xNi_xCu_xCo_xZn_xO$, $x$=0.2), J35($Mg_xNi_xCu_xCo_xZn_xCr_xO$, $x$=0.167) were used to fit above equation. The fitting results showed that the value of $\gamma$ varies with configurational entropy. For binary alloy, $Co_{0.2}Ni_{0.8}O$ has the highest $\gamma$ value of 1.43 while $Zn_{0.4}Mg_{0.6}O$ has the lowest one 0.84, between which is the value of $Co_{0.25}Ni_{0.75}O$ 1.34. Apparently that the value of fitted Gruneisen's parameter will decrease with configurational entropy. And for high entropy oxide J35 and J14, the best fitting results of $\gamma$ both equal 0. That may be explained as the relaxation of elastic strain. For solid solution alloy like $Co_{0.2}Ni_{0.8}O$, the crystal lattice is constructed mainly through host component NiO, and the doping of CoO will break down intrinsic bonding and introduce localized lattice stress and influence the fitting results of Gruneisen's parameter. But with the increasing of doping content like in $Zn_{0.4}Mg_{0.6}O$, host component MgO will not be crystal skeleton anymore. The original localized MgO bond will be gradually replaced by ZnO, which will connect through entire crystal. This process will free the localized residual stress and subsequently reduce the fitting results. And for high entropy structure where all kinds of principal atoms are randomly distributed in certain crystal sites, instead, strain energy will be less prominent because of the composition homogeneity. Therefore, both the values of Gruneisen's parameter of J14 and J35 high entropy oxides were fitted as zero. What's more, in Abeles's research [18], atomic radius term can be treat as proportional to stiffness term. Based on above analysis, in high entropy material materials, we can directly ignore the elastic strain difference term and total disorder degree expression can be reduced as:

$$\Gamma_{tot} = \sum_{N} x_i \left[ \left( \frac{\Delta M_i}{\bar{M}} \right)^2 + 2 \times \left( \frac{\Delta G_i}{\bar{G}} \right)^2 \right] \quad (4)$$

Atomic mass for a certain kind of element is a fix value and thus mass difference can be easily obtained once we know the composition of high entropy system. Nevertheless, for second term, elastic bonding difference term (or force constant disorder), it's still a problem for model simplification. Firstly, there is not even a clear definition of force constant disorder. Elastic bonding difference term $\frac{\Delta G_i}{\bar{G}}$ in Klement's model is based on simple cubic lattice structure and just consider the nearest interatomic bonds. But in real crystals, especially in some crystal structures with large unit-cell, interatomic bonding is extremely complex. What's more, interatomic force between same elements will change with different chemical environment. Therefore, we need to find a simpler indicator to describe the degree of force constant disorder. Some researchers found that lattice distortion (or atomic displacement) have strong relationship with mechanical and thermal properties of high entropy ceramic[23, 24, 30-33]. That make sense because larger bonding elasticity differences would lead to larger atomic displacement after stress relaxation for reaching force balance. Based on this analysis, root square relationship between lattice disorder and force constant disorder was found, which would be clarified in following chapter.

According to Boltzmann equation with relaxation time approximation, thermal conductivity can be expressed as:

$$\kappa = \frac{1}{NV_0}\sum_{\lambda} Cv_{\lambda} \cdot {v_{\lambda}}^2 \cdot \tau_{\lambda} \tag{5}$$

where $N$ is the number of wave vectors, $V_0$ is the volume of primitive cell, $Cv$ is heat capacity, $v$ is the speed of sound (phonon group velocity), $\tau$ is phonon lifetime and $\lambda$ is individual vibration mode. Combining with equation (1), thermal conductivity from phonon-disorder can be rewritten as:

$$\kappa_{ph-dis} = \frac{4\pi}{N{V_0}^2\Gamma_{tot}}\sum_{\lambda}\frac{Cv_{\lambda} \cdot v_{\lambda}}{q^4(\lambda)} = \frac{1}{b\Gamma_{tot}} \tag{6}$$

If we consider the disorder term to be the dominate factor that affect the thermal conductivity then we can treat the rest part as constant factor $b$ that need to be fitted.

Thermal conductivity doesn't just come from phonon- disorder scattering, but also come from phonon-phonon interaction. Here we consider their contribution as independent ones and ignore their possible interaction part. In same crystal structure system, high entropy composition usually has difference of just one kind of element; considering their average effect, we assume their phonon-phonon scattering part to be an average effect under certain temperature. Therefore, we can simply treat it as constant factor $a$ and total thermal conductivity can be further revised as:

$$\kappa_{tot} = \frac{1}{a + b\Gamma_{tot}} \tag{7}$$

But how to deal with disorder term is still unsolved. In Klement's model, elastic differences comes from 1NN neighbors based on simple cubic crystal. In real high entropy crystals, things are more complicated. Some researchers[7] treat elastic bond differences as average of individual component bulk modulus. To some extent, this treatment reflects the disorder of force constant in the crystal. However, atomic bonding varies from different surrounding environment, even if they are between the same element M-O. Not to speaking of complex structure like pyrochlore, which usually has larger unit-cell and atomic force constant disorder cannot be quantified. Therefore, we are demanding a more general indicator to denote the degree of force constant disorder. Here we use lattice disorder (atomic displacement) to replace the force constant disorder. Most importantly, this lattice disorder is easily obtained that can be realized in DFT calculation. We found that, lattice disorder has square root relationship with force constant disorder and one pre-factor *l* need to be fitted. Therefore, expression (7) of thermal conductivity can be written as:

$$\kappa_{tot} = \frac{1}{a + b(\Gamma_m + 2\Gamma_f)} = \frac{1}{a + b(\Gamma_m + 2l\delta_{latt})} \tag{8}$$

where $\delta_{latt}$ is degree of lattice disorder while *l* is conversion factor from lattice disorder to force constant disorder that need fitting. And for the expression of lattice disorder:

$$\delta_{lattice} = \sqrt{\sum_i x_i \left(\frac{d}{\bar{r}_{M-O}}\right)^2} \tag{9}$$

where $x_i$ is fraction of $i^{th}$ element, *d* is displacement of relaxed atoms away from their original position and $\bar{r}_{M-O}$ denotes the average distance of 1st nearest neighbor M-O bonding. And for mass disorder, consider there might be two different crystal sites that need to be replaced, mixed mass disorder can be expressed as:

$$\Gamma_m = \sqrt{{\Gamma_m}^2(A) + {\Gamma_m}^2(B)} \tag{10}$$

To testify our model, we applied it on two systems, one is rock salt oxides, the thermal conductivity of which comes from the indicator of supercell phonon unfolding method, where phonon scattering rate can be seen as linewidth of dispersion spectra; the other system is pyrochlore structure, which have two doping site, composition design and experimental thermal conductivity data of which come from Wright's research[34] (data shown in Table 2).

# 3. Results and discussion

## 3.1 lattice disorder and force constant disorder

Lattice disorder is the common character found in disorder system including medium and high entropy materials [35-39]. And this character can be regarded as the source of unique properties such as high strength, high toughness, and low thermal conductivity. Fig. 1(a) is the schematic of one atomic layer of high entropy rock salt oxide structure after relaxation. The red balls represent oxygen atoms while other colored balls denote randomly distributed cations. Rocksalt oxides structure has two crystal sites, 4a and 4b. All the oxygen atoms take 4b sites while one element take 4a position. Under ideal crystal symmetry, all the atoms are supposed to keep staying at cross points of lattice shown in Fig.1 (a) of dash lines. However, high entropy oxides requires original 4a sites to be randomly replaced by multiple cations (like $Mg^{2+}$, $Zn^{2+}$, $Co^{2+}$, $Cu^{2+}$, $Ni^{2+}$, etc.), which inevitably breaks the symmetry of lattice and change the balance position of every atoms. The new positions slightly deviate from the ideal ones and thus long range order and can still be identified by experiments like X-ray diffraction pattern[7]. Fig. 1(a) presents ideal position with black dots while relaxed atomic positions with white dots. To reveal the lattice distortion, only the lattice wireframe was extracted and shown in Fig. 1 (b), from which we can tell all the unite cell lattice is distorted and not be perfect cubic shape anymore. For different unit cell, the degree of distortion looks different. Some are ladder-shaped, some are diamond-shaped and some are just in random quadrilateral shape. The origin of the distortion is the inhomogeneous atomic interaction, which break up the intrinsic force balance and would build a new one. This force readjustment leads to atomic displacement and subsequentially causes lattice distortion. Some study believe the ionic size difference plays a key role in atomic relaxation and influence the macro performance[9, 34]. However, in real materials, the ion size is not fixed and change from different environment. Therefore, lattice distortion should be the one that we can make use of. Importantly, lattice distortion is an easily obtained indicator through ab-initio calculation. As we introduced in method part, force constant disorder is the essence that influence the phonon scattering and lattice distortion is just the appearance. Now the problem lies on whether we can successfully find or define the connection between lattice distortion and force constant disorder.

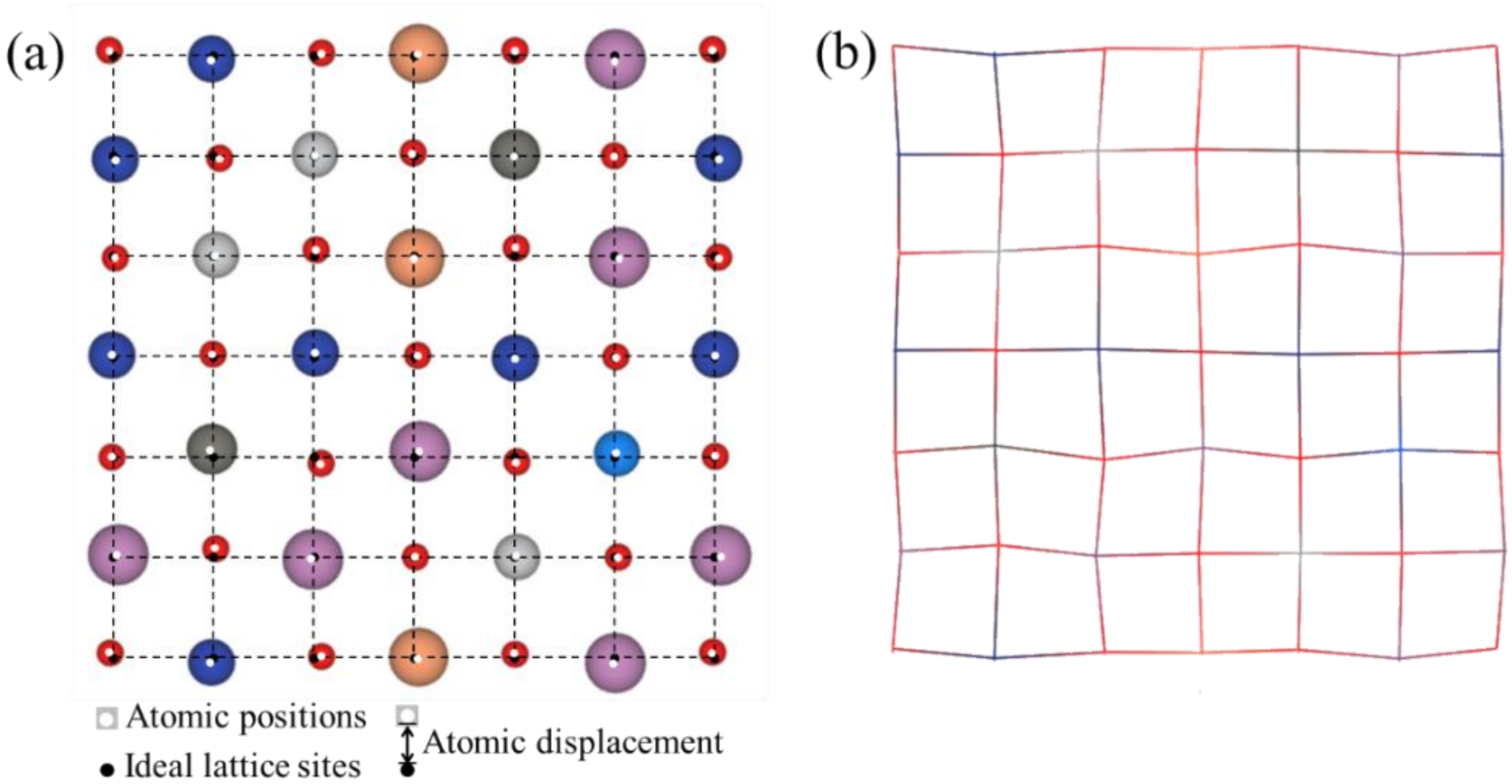

Figure 1 (a) One (100) crystal face of high entropy rock salt oxides, (b) Distorted lattice frame of (a).

Firstly, we constructed supercell of 19 kinds of medium and high entropy rocksalt oxides through SQS method, composition of which is inspired by Braun's research[7] and displayed in Table 1, to explore the relationship between lattice disorder and force constant disorder. To ensure the supercell to be as larger as possible to present high entropy effect, meanwhile considering our high-performance cluster calculation capability, we created supercell that includes 216-250 atoms for different composition. After volume optimization and atomic relaxation, and comparing the position difference before and after relaxation, we can collect the atomic displacement of every atoms. Fig 2 displays atomic displacements distribution of partial compositions, which gradually increase the cations element from 3 to 10, with calculated degree of lattice disorder shown at top right corner. All the lattice disorder degree were calculated and present in corresponding composition. Overall, the lattice disorder will increase with number of cation elements. But this is not strict, lattice disorder of 5 dopants (Fig. 2(b)) is smaller than that of 4 dopants (Fig. 2(c)). 5 cations and below, displacement of oxygen anions is much larger than other metal elements; above 6 cations, largest displacement of metal element can be in the same level as oxygen. There are two times of sharply increasement of degree of lattice disorder. First is when fourth cation $Cu^{2+}$ was added, where lattice disorder increase nearly 6 times from 0.01065 to 0.05951. The second time is the adding of sixth cation of Sn+, lattice disorder degree increase from 0.04820 to 0.15848 (three times). The sixth cation change the displacement distribution of $Cu^{2+}$ ion, which corresponds the suddenly decrease of thermal conductivity in Braun's research[7]. It seems that when cation type exceed 10, the lattice disorder will keep steady and hard to increase like when the cation type is small. To explore the essence of it, further investigation on atomic bonding is necessary.

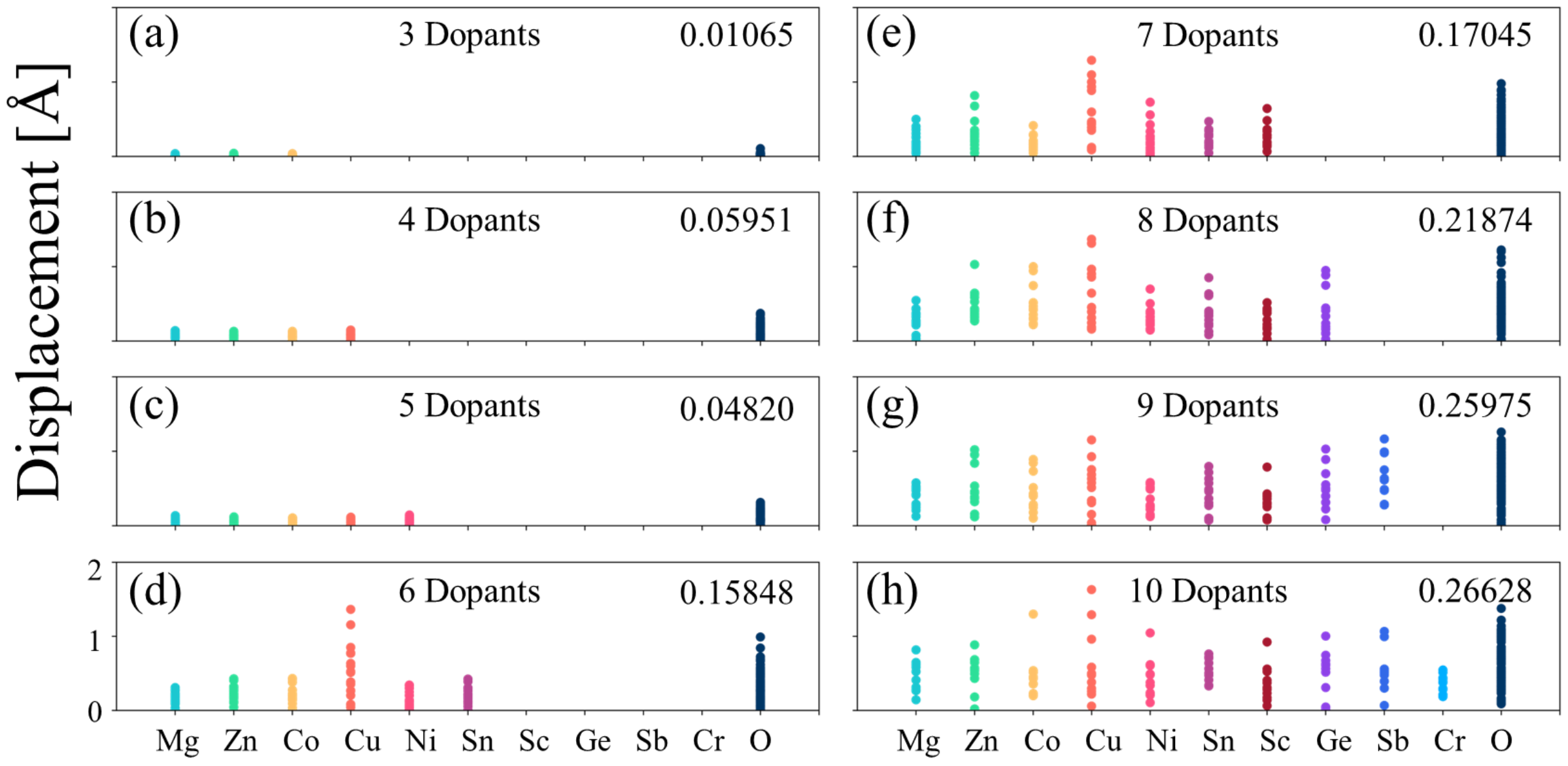

Figure 2 Atom displacements with the increasement of cation elements with lattice disorder degree shown at top right corner.

Based on finite displacement method (through *phonopy* python package[40]), we calculated the $2^{nd}$ order force constant (FC) of all the medium and high entropy rock salt oxides and collected every $2^{nd}$ order FC between first nearest neighbor (1NN) M-O bond. $2^{nd}$ order FC is a $2^{nd}$ rank tensor which includes 9 values, 6 non-diagonal elements ($\Phi_{xy}$, $\Phi_{yx}$, $\Phi_{yz}$, $\Phi_{zy}$, $\Phi_{xz}$, $\Phi_{zx}$) and 3 diagonal elements ($\Phi_{xx}$, $\Phi_{yy}$, $\Phi_{zz}$). To simplify the model, we just consider the diagonal elements $\Phi_{nn}$ as effective data, the value of which is much larger than non-diagonal elements. Here we used equation:

$$\Gamma_f = \frac{1}{3N} \sum_{\substack{i=N \\ j=xx,yy,zz}} \left(1 - \frac{\Phi_{i,j}}{\bar{\Phi}_{nn}}\right)^2 \qquad (11)$$

to calculate the force constant disorder, where N is the number of 1NN M-O bonding, $\bar{\Phi}_{nn}$ is the average of all the $\Phi_{nn}$ components, $\Phi_{i,j}$ is individual tensor elements that includes *xx*, *yy* and *zz* component of all the 1NN M-O bonding. The diagonal element $\Phi_{xx}$, $\Phi_{yy}$, $\Phi_{zz}$ correspond to two different bonding, when atomic direction is the same as subscript, this force can be regard as normal stress while other two can be regard as shear stress, both of which contributes to atomic vibration along the interatomic direction. Ideally, normal stress and shear stress need to be considered separately because they are responsible for different vibration mechanism. However, in simple crystal like rocksalt structure, this assumption make some sense. And importantly, the structure of rocksalt oxide is as simple cubic crystal, the force constant disorder of which to some degree can be reflected from equation (11). The following analysis will approve that in high entropy ceramic, the differences between normal stress and shear stress is not as distinct as in perfect crystals.

Figure 3 is the FC distribution histogram in 3-cations medium entropy rocksalt ceramic, MgZnCoO and MgZnCuO (Simplification format of $(Mg_{1/3}Zn_{1/3}Co_{1/3})O$ and $(Mg_{1/3}Zn_{1/3}Cu_{1/3})O$). For composition of MgZnCoO, shown in Fig. 3 (a), we can see that the force distribution doesn't follow any rules, but seems have three peaks, the average of force constant is -1.0594 eV/m while force constant disorder is 0.1109. If we separate the force constant according to different metal elements, we can obtain Fig. 3 (b) - (d). For each kind of M-O bonding, the force constant distribution have two kind of peak, one is centered at smaller value, from -1 eV/m to -2 eV/m, which presents normal stress between two atoms, while the other peak corresponds shear stress part, which is on the right side of -1 eV/m and more concentrated. But things are different in composition of MgZnCuO. Fig. 3 (e) shows total force constant distribution histogram. Compared with MgZnCoO, the total force constant distribution looks like one widely distributed peak. Although with few counting numbers, the force constant has some distribution smaller than -2 eV/m, which means in this composition, lattice disorder is severer than that in MgZnCoO. From average force constant, the value of MgZnCuO is -0.9339 eV/m, which is similar of MgZnCoO. However, the force constant disorder of MgZnCuO is 1.1544, much larger than the other one. To investigate the mechanism behind it, individual M-O force constant is separated and shown in Fig.3 (f) - (h). For Mg-O bonding, force distribution is totally different. Comparing to Fig.3 (b), the whole peak of force constant distribution slightly moves to right, and distinction between normal and shear stress is not apparent anymore. Therefore, put normal and shear stress together to analyze is reasonable. For Zn-O bonding, distribution change is same as what happens in Mg-O. But for the 3$^{rd}$ element Cu, its force constant distribution is different as other bondings. The distribution peak of Cu-O force constant is wider and more flat than other elements. This phenomenon happens in all the composition that contains Cu when the dopant number is below 6. It means that Cu plays an important role for broadening force constant distribution and enhance the force constant disorder degree. This triggering effect of Cu might come from Jahn-Teller effect, which usually happened in Cu(II). Jahn-Teller effect is a mechanism of spontaneous symmetry breaking that usually happens in transition metal elements of octahedral complexes[41-43]. The $d^9$ electron configuration of the ion will give three electron to two degenerate $e_g$ orbital, leading to distortion of complex along one fourfold axis to reach energy minimization. This theory explains larger lattice distortion after adding of Cu element. And once the lattice distortion happened, interatomic direction will be further away from crystal orthogonal basis direction and lead to larger force constant disorder. Under large distortion, shear stress will not be differentiable from normal stress and they will act together on the surrounding atoms.

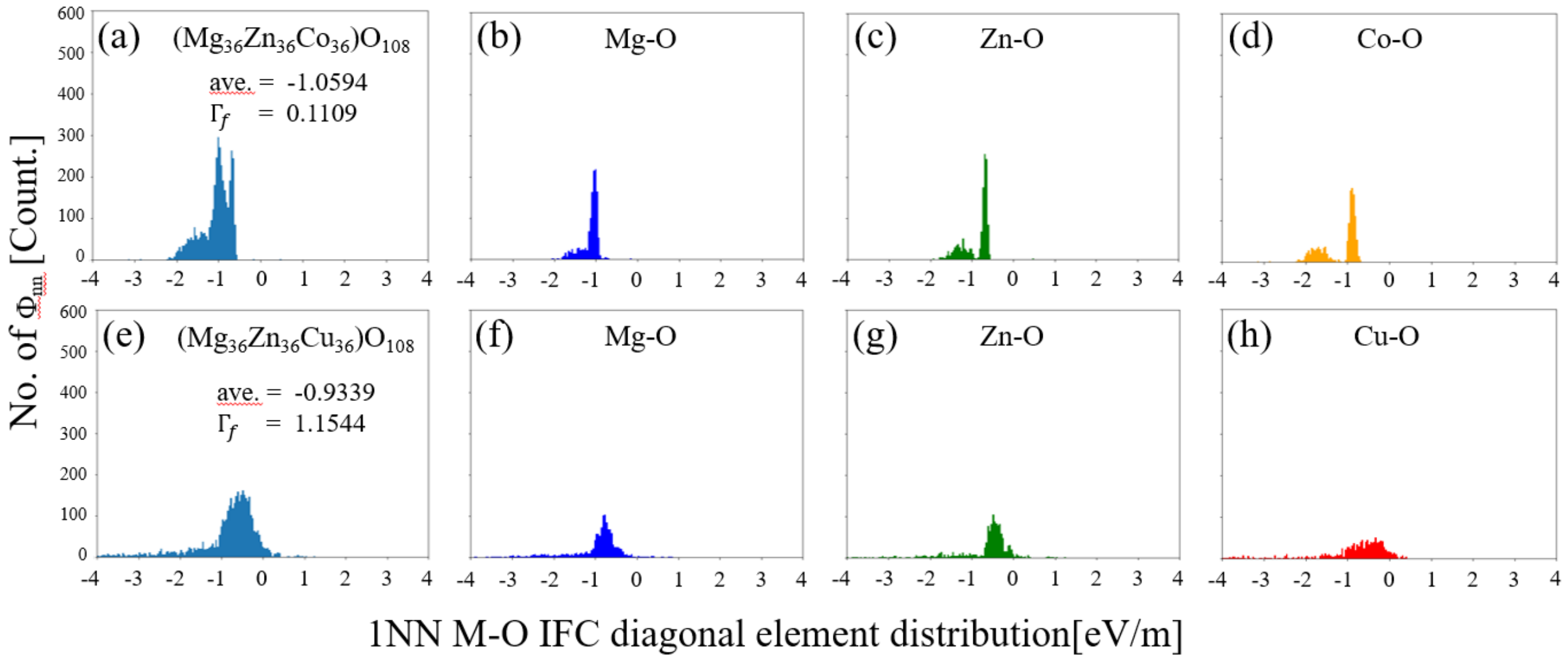

Figure 3 1NN IFC distribution of diagonal elements $\Phi_{nn}$ in MgZnNiO and MgZnCuO. (a), (e) total bonding histogram. (b-d), (f-h) individual M-O bonding histogram.

Another change need to notice is that when cation type increase from 5 to 6, both the lattice disorder and force constant disorder will be increased by factor about 3. To explore the mechanism behind it, $(Mg_{25}Zn_{25}Co_{25}Cu_{25}Ni_{25})O_{125}$ and $(Mg_{18}Zn_{18}Co_{18}Cu_{18}Ni_{18}Sb_{18})O_{108}$ were selected and their force constant disorder were calculated through distribution shown in Fig. 4. The only difference of this two composition is the adding of Sb element as the 6th cation type. Their average force constant were -1.0231 eV/m and -1.2374 eV/m respectively while their degree of disorder were 0.6622 and 2.1636. This phenomenon leads to further decrease of thermal conductivity, which have been observed in experiment results[7]. From total distribution histogram, the distribution profile of 6 cations oxide is lower than that of 5 cations, which means more widely distributed. For 6 cations composition, some of the force constant moved to positive range, which denotes part of the atomic distance surpass the local minimum of interatomic force-distance curve, which is much larger than the balance distance. And if we look at the individual metal oxygen bonds in Fig.4 (b-f) and (h-m), we can find the differences mainly come from Cu and Sb. For other elements' bonded with surrounding oxygen like Mg, Zn, Co and Ni, the position of distribution peak doesn't change much, but the peak height decreases a little. But for bonding between Cu and surrounding O atoms, the change is prominent. Firstly, peak position moved from negative value right to nearly around zero, which corresponds to minimum point of interatomic force-distance curve. And the shape of Cu-O force constant distribution become narrower, which is as that of Mg, Zn, Co and Ni. The drastic change of this Cu-O force constant distribution is believed to stem from the addition of sixth cation element Sb. The outer electron configuration of Sb might have impact on Cu electron configuration, which could reverse the influence from Jahn-Teller effect and recover the Cu-O bonding back to octahedral complex. At the same time, Sb-O force constant distribution itself is extreme widely distributed, nearly in flat shape ranging from -2 eV/m to 0 eV/m. This highly distorted complex structure will undoubtedly increase the total force constant disorder in this crystal structure.

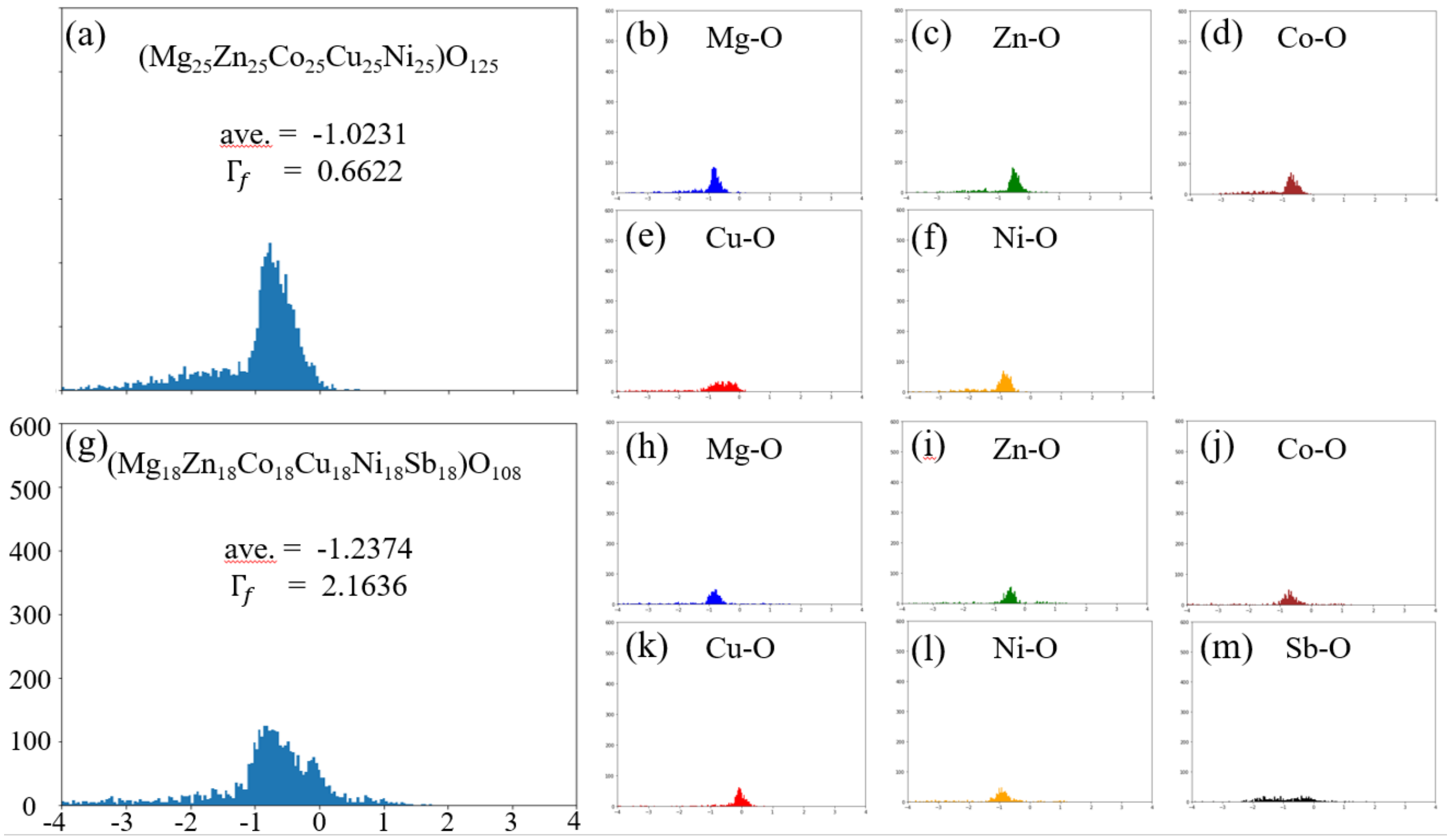

Figure 4 1NN IFC distribution of diagonal elements Φnn in MgZnCoCuNiO and $MgZnCoCuNiSbO_{108}$. (a), (g) total bonding histogram. (b-f), (h-m) individual M-O bonding histogram.

Now we have lattice disorder and force constant disorder in rocksalt medium and high entropy oxides. Theoretically, when there is no lattice disorder, there will be no force constant disorder. If we look at the data distribution of lattice and force constant disorder show in Fig. 5, their relationship is not like linear one. Thus firstly, we tried root square relationship $\Gamma_f = l \times \delta_{lattice}^{0.5}$ where $l$ is a constant factor that needed fitting. We used non-linear least squares to fit the curve. The fitting result shows that the coefficient of determination reached 0.901, which is high enough to be adopted in our future study; and pre-factor $l$ equals 4.325, which can be regard as the conversion factor from lattice to force constant disorder. Despite that in different crystal structure, especially for complex structure oxide, the expression of force constant disorder degree might not be same as equation (11), the root square relationship between force and lattice disorder should be meaningful. Therefore, for other structure, like pyrochlore medium/high entropy oxides, we can still use $l$ to represent the conversion factor from lattice to force constant disorder. And in the following chapters, we would try to build the prediction model and examineed on two different structure rocksalt oxides and pyrochlore oxides.

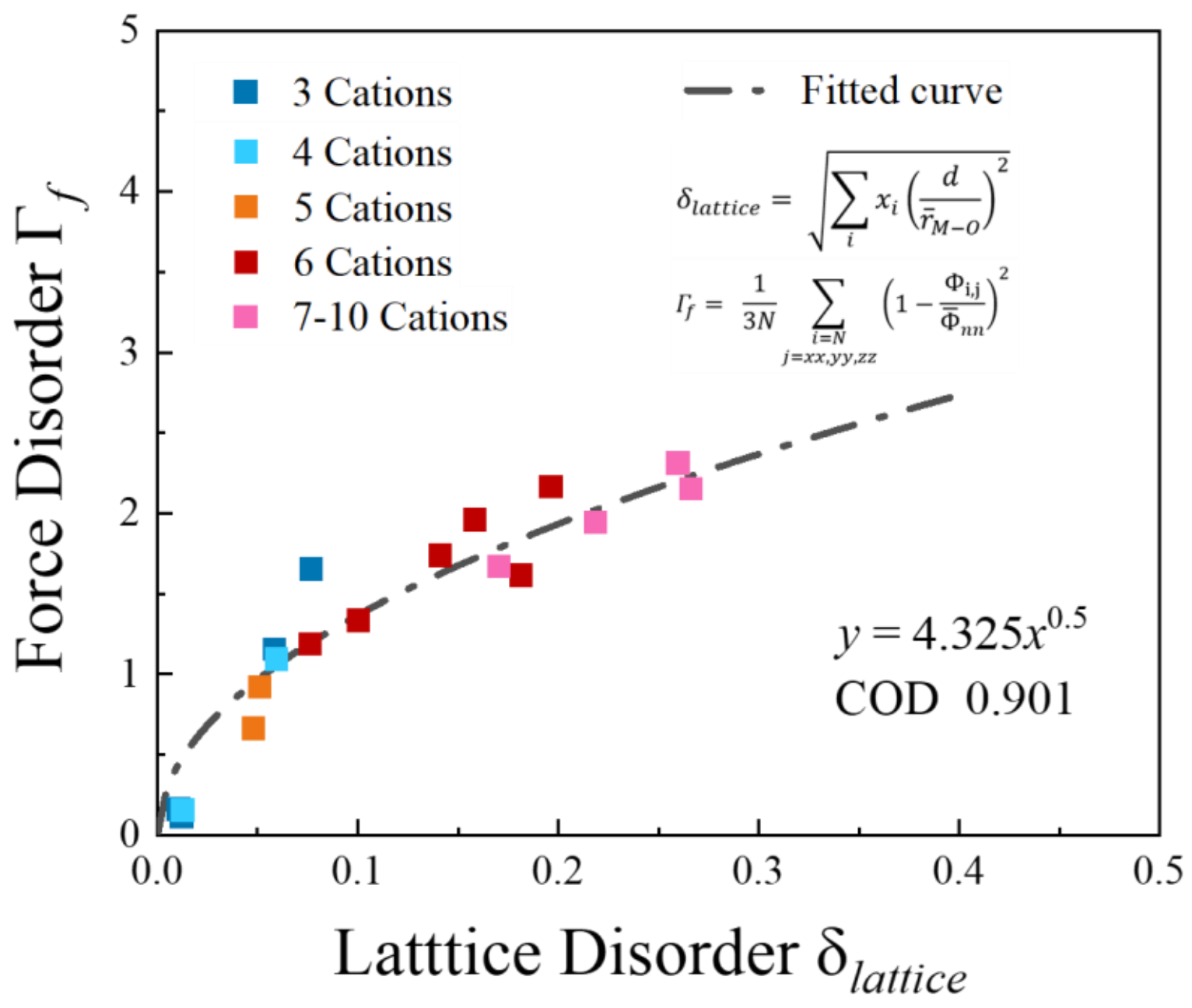

Figure 5 Square root relationship between force constant disorder and lattice disorder.

## 3.2 Application of models on rock salt structure and SPU models

Once we built root square relationship between force constant disorder and lattice disorder, our prediction model can be built and tested in rocksalt oxide structure first. The whole contribution of thermal conductivity can be regarded as expression (8). There are 3 parameters needed to be fitted, $a$ is the thermal conductivity contribution from phonon-phonon interaction while $b$ can be seen as the constant factor that connect disorder degree to thermal conductivity.

And for parameter *l*, it can be treated as the convert factor from lattice disorder to force constant disorder.

Firstly, we applied our model on rocksalt oxide structure. Because we have already obtained conversion factor *l* in previous chapter which equals 4.325, only *a* and *b* need to be fitted. In our previous study, phonon lifetime of phonon-disorder can be approximated through supercell phonon unfolding (SPU) method and thus the thermal conductivity from which can be calculated. Therefore, through this method, we calculated the thermal conductivity indicator of all the designed composition and shown in Table. 1. Here we used the results from SPU as y coordinate data to do the fitting, and the results are shown as Fig 6. The fitting result shows parameter *a* equals to 0 while *b* equals 0.027. The value of *a* corresponds to the contribution from phonon-phonon interaction, and the data of the y coordinate is from SPU method, which doesn't combine the contribution from phonon-phonon scattering part. This means the fitting results of *a* matches well with our thermal conductivity data setting and prove that our model works perfect in rocksalt oxide structure. The fitting curve shows a reciprocal relationship between disorder degree and thermal conductivity indicator, which means that with the increase of disorder degree, the thermal conductivity will decrease gradually. However, the degree of disorder will not increase infinitely, which will be limited by crystal structure and phase stability. Therefore, theoretically, there might exist the lowest value of thermal conductivity in high entropy system but need further study. And in Fig. 6, we also labeled the composition of different cation number with different color, from which we can easily tell that disorder degree can be a better indicator that cation number to denote the property of thermal conductivity.

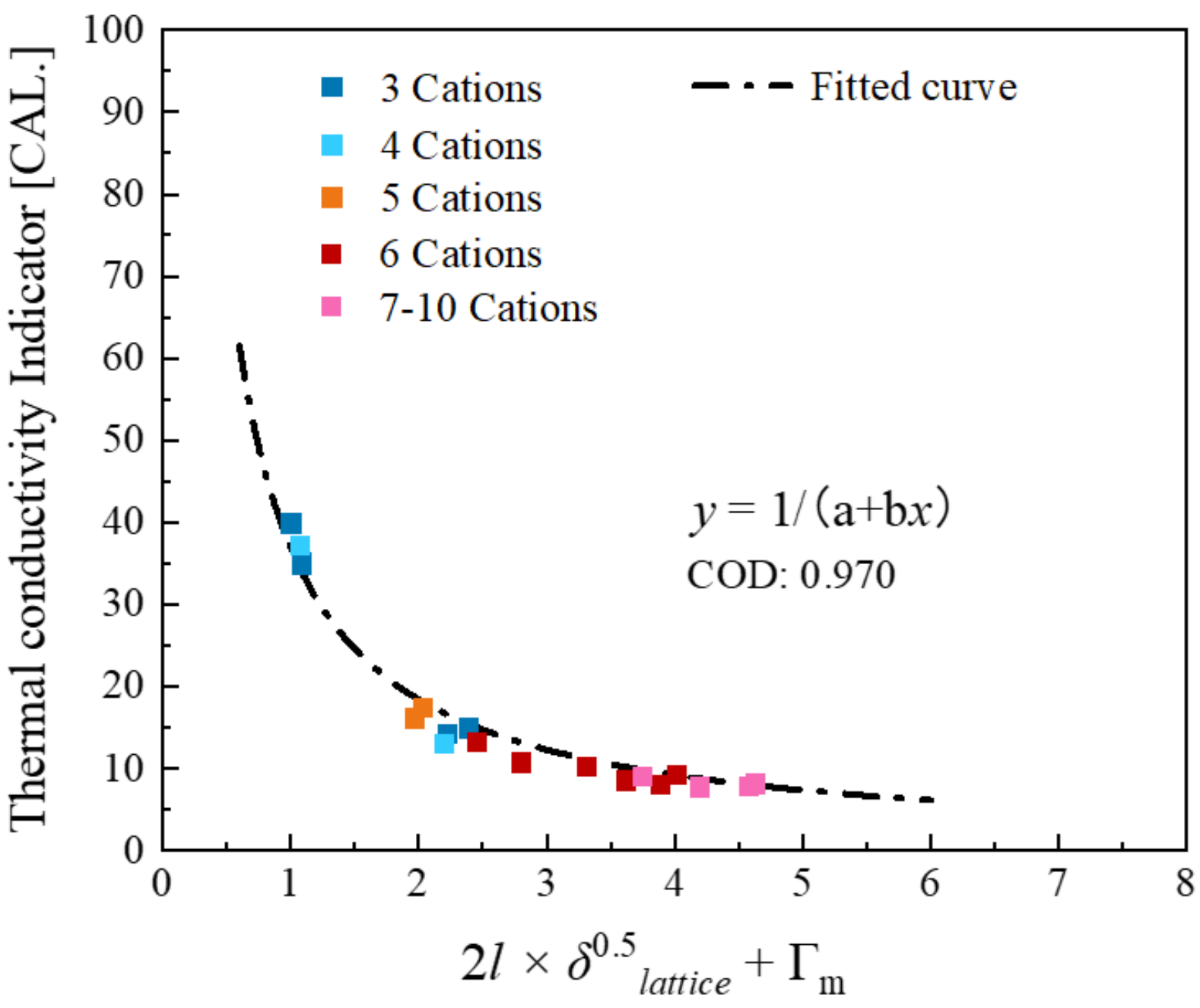

Figure 6 Model fitting on medium and high entropy rocksalt oxides.

### 3.3 Application on experimental results of pyrochlore structure

To testify the general applicability, we applied our model on medium and high entropy pyrochlores, the composition design and room temperature experimental thermal conductivities of which come from Wright's research[34] and the data are listed in Table 2. The structure of pyrochlore is $A_2B_2O_7$ thus there are two possible atomic sites for doping. For A position, which present 3+ valence, the dopant elements are rare earth lanthanides: La, Sm, Gd, Eu, Pr, Nd, Yb. And for B position with 4+ valence, the dopant elements are Ti, Sn, Hf, Zr. From Wright's results[34], $(Sm_{1/3}Eu_{1/3}Gd_{1/3})_2Ti_2O_7$ has the highest thermal conductivity of 2.88 $W{\cdot}m^{-1}{\cdot}K^{-1}$ while $(Sm_{1/4}Eu_{1/4}Gd_{1/4}Yb_{1/4})_2(Ti_{1/4}Sn_{1/4}Hf_{1/4}Zr_{1/4})_2O_7$ owns the lowest value of 1.45 $W{\cdot}m^{-1}{\cdot}K^{-1}$. The degree of lattice disorder and mass disorder were calculated from equation (9) and (10). The results of model applying on medium and high entropy pyrochlore oxides are present in Fig. 7. Under best fitting, when $l$ equals to 0.197, a equals 0.314 and b equals 1.102, the coefficient of determination reaches the highest value of 0.922, which is also at a high level and enough to prove the feasibility of our model. It's noticed that parameter $a$ equals 0.314. Because the factor $a$ reflect the phonon-phonon scattering contribution, we can calculate the thermal conductivity as the reciprocal value of $a$, which is about 3.190 $W{\cdot}m^{-1}{\cdot}K^{-1}$. This value is slight higher than experimental result of single component $La_2Zr_2O_7$ [34], which is the structure base of all the high entropy pyrochlore compositions in this research. From equation (7), thermal conductivity has two contributions, phonon-phonon scattering (*a*) and phono-disorder scattering (*b*x), and their contribution can be expressed as reciprocal value 1/*a* and 1/(*b*x) respectively, which were also plotted in Fig.7 as short red dot line and long blue dot line. We can see that when degree of disorder is small, phonon-phonon scattering take the dominant contribution. And with the increase of disorder degree, the contribution from phonon-disorder increased and comparable with that from phonon-phonon part. Further increase the disorder degree, the total thermal conductivity will depend on the phonon-disorder contribution. When inherent phonon-phonon thermal conductivity is high, the high entropy effect to reduce thermal conductivity will be prominent, which usually happens in simple crystals like rocksalt structure. In Braun's research[7], one element adding would even lead to nearly half reduction of thermal conductivity in rocksalt structure. In this paper, we assume that their contributions as independent ones and regard phonon-phonon contribution as constant value under certain temperature. However, the real situation might be more complicated as they have coupled effect in this two mechanisms and thus need further theoretical study.

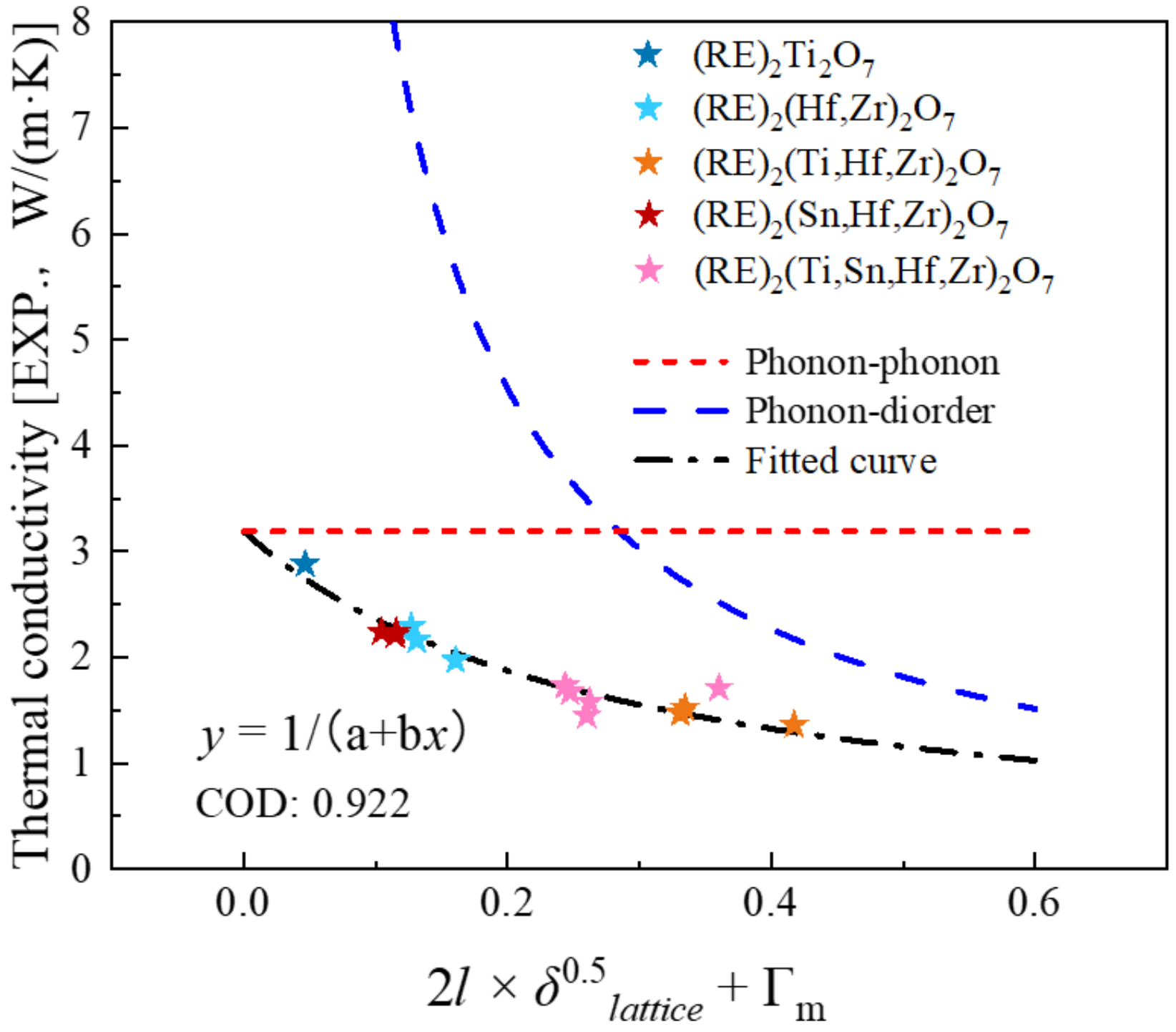

Figure 7 Model fitting on medium and high entropy pyrochlore oxides.

# 4. Summary

In this paper, we constructed a quantitative prediction model for thermal conductivity of medium and high entropy oxides materials. This model combine the effects from both phonon-phonon and phonon-disorder scattering part. In phonon-disorder expression, we used easily obtained and more general lattice disorder indicator to replace complicated force constant disorder. Our model were examined in two different medium/high entropy system, rocksalt oxide and pyrochlore oxide. The results of our prediction model have every good alignment with calculational and experimental thermal conductivity; and the coefficient of determination in two system reached 0.970 and 0.922, which is high enough to prove the feasibility of it. This model provide a high efficiency way to predict thermal conductivity of medium/high entropy ceramic system and accelerate the development of next-generation ceramic materials with low thermal conductivity.

# 5. Appendix

Table 1 Composition design, relative disorder degree and thermal conductivity information of medium/high entropy rocksalt oxides.

| Num. | Serial | Dopants No. | Composition | Lattice Disorder | Mass Disorder | EXP. TC ($W \cdot m^{-1} \cdot K^{-1}$) | SPU Indicator ($W \cdot m^{-1} \cdot K^{-1}$) |
|---|---|---|---|---|---|---|---|
| 1 | D3 | 3 | $(Mg_{36}Co_{36}Ni_{36})O_{108}$ | 0.0107 | 0.1182 | N/A | 39.803 |
| 2 | D3 | 3 | $(Mg_{36}Zn_{36}Co_{36})O_{108}$ | 0.0123 | 0.1326 | N/A | 34.736 |
| 3 | D3 | 3 | $(Mg_{36}Zn_{36}Cu_{36})O_{108}$ | 0.0585 | 0.1376 | N/A | 14.140 |
| 4 | D3 | 3 | $(Mg_{36}Zn_{36}Ni_{36})O_{108}$ | 0.0123 | 0.1325 | N/A | 35.144 |
| 5 | D3 | 3 | $(Zn_{36}Cu_{36}Ni_{36})O_{108}$ | 0.0769 | 0.0020 | N/A | 14.848 |
| 6 | D4 | 4 | $(Mg_{27}Zn_{27}Co_{27}Cu_{27})O_{108}$ | 0.0595 | 0.0998 | N/A | 12.932 |
| 7 | D4 | 4 | $(Mg_{27}Zn_{27}Co_{27}Ni_{27})O_{108}$ | 0.0129 | 0.0967 | N/A | 37.172 |
| 8 | J14 | 5 | $(Mg_{25}Zn_{25}Co_{25}Cu_{25}Ni_{25})O_{125}$ | 0.0482 | 0.0783 | 2.95 | 16.026 |
| 9 | J30_xCu | 5 | $(Mg_{25}Zn_{25}Co_{25}Ni_{25}Sc_{25})O_{125}$ | 0.0513 | 0.0846 | N/A | 17.307 |
| 10 | J30 | 6 | $(Mg_{18}Zn_{18}Co_{18}Cu_{18}Ni_{18}Sc_{18})O_{108}$ | 0.0762 | 0.0734 | 1.68 | 13.149 |
| 11 | J31 | 6 | $(Mg_{18}Zn_{18}Co_{18}Cu_{18}Ni_{18}Sb_{18})O_{108}$ | 0.1965 | 0.1930 | 1.41 | 9.157 |
| 12 | J34 | 6 | $(Mg_{18}Zn_{18}Co_{18}Cu_{18}Ni_{18}Sn_{18})O_{108}$ | 0.1585 | 0.1826 | 1.44 | 8.475 |
| 13 | J35 | 6 | $(Mg_{18}Zn_{18}Co_{18}Cu_{18}Ni_{18}Cr_{18})O_{108}$ | 0.1004 | 0.0664 | 1.64 | 10.648 |
| 14 | J36 | 6 | $(Mg_{18}Zn_{18}Co_{18}Cu_{18}Ni_{18}Ge_{18})O_{108}$ | 0.1412 | 0.0729 | 1.60 | 10.117 |
| 15 | J30_Cu-Sn | 6 | $(Mg_{18}Zn_{18}Co_{18}Sn_{18}Ni_{18}Sc_{18})O_{108}$ | 0.1816 | 0.2162 | N/A | 7.993 |
| 16 | D7 | 7 | $(Mg_{16}Sc_{15}Zn_{16}Co_{15}Cu_{15}Ni_{16}Sn_{15})O_{108}$ | 0.1705 | 0.1843 | N/A | 8.995 |
| 17 | D8 | 8 | $(Mg_{14}Sc_{13}Zn_{14}Co_{14}Cu_{13}Ni_{14}Sn_{13}Ge_{13})O_{108}$ | 0.2187 | 0.1572 | N/A | 7.664 |
| 18 | D9 | 9 | $(Mg_{12}Sc_{12}Zn_{12}Co_{12}Cu_{12}Ni_{12}Sn_{12}Ge_{12}Sb_{12})O_{108}$ | 0.2598 | 0.1840 | N/A | 7.702 |
| 19 | D10 | 10 | $(Mg_{10}Sc_{11}Zn_{10}Cr_{11}Co_{11}Cu_{11}Ni_{11}Sn_{11}Ge_{11}Sb_{11})O_{108}$ | 0.2663 | 0.1778 | N/A | 8.027 |

Table 2 Composition design, relative disorder degree and thermal conductivity information of medium/high entropy pyrochlore oxides.

| **Serial** | **Composition** | **Lattice disorder** | **Mass disorder** | **EXP. TC ( $W\cdot m^{-1}\cdot K^{-1}$)** |
|---|---|---|---|---|
| P1 | $La_2(Hf_{1/2}Zr_{1/2})_2O_7$ | 0.0069 | 0.1047 | 2.29 |
| P4 | $Sm_2(Sn_{1/4}Ti_{1/4}Hf_{1/4}Zr_{1/4})_2O_7$ | 0.0423 | 0.1886 | 1.73 |
| P5 | $Gd_2(Sn_{1/4}Ti_{1/4}Hf_{1/4}Zr_{1/4})_2O_7$ | 0.0760 | 0.1886 | 1.58 |
| P6 | $(Sm_{1/2}Gd_{1/2})_2(Ti_{1/3}Hf_{1/3}Zr_{1/3})_2O_7$ | 0.0644 | 0.2633 | 1.47 |
| P7 | $(Eu_{1/2}Gd_{1/2})_2(Ti_{1/3}Hf_{1/3}Zr_{1/3})_2O_7$ | 0.0710 | 0.2633 | 1.52 |
| P8 | $(La_{1/2}Pr_{1/2})_2(Sn_{1/3}Hf_{1/3}Zr_{1/3})_2O_7$ | 0.0088 | 0.0792 | 2.24 |
| P9 | $(Eu_{1/2}Gd_{1/2})_2(Sn_{1/3}Hf_{1/3}Zr_{1/3})_2O_7$ | 0.0185 | 0.0792 | 2.24 |
| P10 | $(La_{1/3}Pr_{1/3}Nd_{1/3})_2(Hf_{1/2}Zr_{1/2})_2O_7$ | 0.0097 | 0.1047 | 2.16 |
| P11 | $(Sm_{1/3}Eu_{1/3}Gd_{1/3})_2(Hf_{1/2}Zr_{1/2})_2O_7$ | 0.0440 | 0.1047 | 1.97 |
| P20-1 | $(Sm_{1/3}Eu_{1/3}Gd_{1/3})_2(Sn_{1/3}Hf_{1/3}Zr_{1/3})_2O_7$ | 0.0176 | 0.0792 | 2.21 |
| P20 | $(Sm_{1/3}Eu_{1/3}Gd_{1/3})_2(Ti_{1/4}Sn_{1/4}Hf_{1/4}Zr_{1/4})_2O_7$ | 0.0486 | 0.1886 | 1.67 |
| P20+Yb | $(Sm_{1/4}Eu_{1/4}Gd_{1/4}Yb_{1/4})_2(Ti_{1/4}Sn_{1/4}Hf_{1/4}Zr_{1/4})_2O_7$ | 0.0711 | 0.1887 | 1.45 |
| P20-2 | $(Sm_{1/3}Eu_{1/3}Gd_{1/3})_2(Ti_{1/2}Sn_{1/6}Hf_{1/6}Zr_{1/6})_2O_7$ | 0.0574 | 0.2962 | 1.71 |
| P20-4 | $(Sm_{1/3}Eu_{1/3}Gd_{1/3})_2Ti_2O_7$ | 0.0298 | 0.0004 | 2.88 |
| P20-5 | $(Sm_{1/4}Eu_{1/4}Gd_{1/4}Yb_{1/4})_2(Ti_{1/2}Hf_{1/4}Zr_{1/4})_2O_7$ | 0.0824 | 0.3407 | 1.36 |

# 6. Acknowledgements

The author acknowledges financial support by National Science and Engineering Research Council of Canada (Grant #: NSERC RGPIN-2023-03628), and the Digital Research Alliance of Canada for providing computing resources.